\documentclass[11pt]{article}

\usepackage[T1]{fontenc}
\usepackage[utf8]{inputenc}
\usepackage[english]{babel}
\usepackage{lmodern}
\usepackage[a4paper,margin=2.4cm]{geometry}
\usepackage{graphicx}
\usepackage{amsmath,amssymb,bm}
\usepackage{booktabs}
\usepackage{caption}
\usepackage{subcaption}
\usepackage{authblk}
\usepackage{microtype}
\usepackage[hidelinks]{hyperref}

\graphicspath{{./figures/}}

\newcommand{\qtip}{q-TIP4P/F}
\newcommand{\varpro}{variable-projection}
\newcommand{\vect}[1]{\mathbf{#1}}
\newcommand{\mat}[1]{\mathbf{#1}}

\title{Separable Force Matching of PBE0 Hybrid-Functional Reference Forces for Path-Integral Simulations of Liquid Water}

\author[1,2]{Jan Kessler}
\author[1]{Thomas Spura}
\author[1]{Kristof Karhan}
\author[3,4,5]{Thomas D. K\"uhne\thanks{CONTACT Thomas D. K\"uhne. Email: \texttt{tkuehne@cp2k.org}}}

\affil[1]{Department of Chemistry, University of Paderborn, Warburger Str. 100, D-33098 Paderborn, Germany}
\affil[2]{Department of Physics, Johannes Gutenberg University Mainz, Staudingerweg 7, D-55128 Mainz, Germany}
\affil[3]{Center for Advanced Systems Understanding (CASUS), Conrad-Schiedt-Strasse 20, 02826 G\"orlitz, Germany}
\affil[4]{Helmholtz-Zentrum Dresden-Rossendorf, Bautzner Landstrasse 400, 01328 Dresden, Germany}
\affil[5]{Institute of Artificial Intelligence, Technische Universit\"at Dresden, Helmholtzstrasse 10, 01069 Dresden, Germany}

\date{}

\begin{document}

\maketitle

\begin{abstract}
Force-matched water models provide a practical route from first-principles reference data to long classical and path-integral molecular simulations.  Previous flexible four-site potentials in the spirit of \qtip{} showed that fitting analytic models to density-functional-theory forces can reproduce key structural features of liquid water while retaining the efficiency required for quantum-nuclear sampling.  Here we introduce two refinements aimed at making this strategy more accurate and more reproducible for molecular simulation.  First, the production fit is based on PBE0 hybrid-functional reference forces and therefore includes the Hartree--Fock exact-exchange contribution in the electronic-structure target.  Second, the parametrization is formulated as a separable nonlinear least-squares problem in which all linear force-field amplitudes are eliminated analytically for every trial set of nonlinear shape parameters.  The resulting force-matching protocol lowers the dimension of the nonlinear search, reduces compensation between heterogeneous parameters, and enables a controlled comparison of Lennard-Jones and Buckingham oxygen--oxygen repulsion terms.  Applied to CP2K reference forces for liquid water, the projected optimization reproduces target force distributions and yields radial distribution functions close to first-principles and neutron-scattering benchmarks.  The Buckingham representation gives a more flexible short-range repulsive wall than a purely Lennard-Jones form, and the final flexible model remains stable in path-integral simulations.  The PBE0-LJ reference fit and the final PBE0-oogam parameter set are reported explicitly and were tested in classical and path-integral trajectories with the \texttt{RPMD\_Mainz} simulation code.  The method provides a transparent workflow for deriving simulation-ready water potentials from accurate hybrid density-functional reference forces.
\end{abstract}


\section{Introduction}

Liquid water remains a demanding target for molecular dynamics (MD) simulation because its structure and dynamics reflect a delicate balance of electrostatics, Pauli repulsion, dispersion, polarization, intramolecular flexibility, and nuclear quantum effects.  Classical empirical water models have been remarkably successful, especially when they are tuned directly to selected experimental observables \cite{Abascal2005,Vega2011}.  However, empirical accuracy can come at the price of an ambiguous connection to an underlying electronic-structure reference, and transfer outside the thermodynamic or structural range used in the parametrization is not guaranteed.  Conversely, first-principles MD avoids an empirical intermolecular potential but is too expensive for routine path-integral molecular dynamics (PIMD), where each atom is represented by an extended ring polymer and the cost of force evaluation is multiplied by the number of beads.  Direct ab initio PIMD is the formally cleanest way to combine electronic-structure forces with explicit nuclear quantum fluctuations.  Its foundations were laid in the pioneering work of Marx and Parrinello and in the efficient path-integral Car--Parrinello algorithms of Tuckerman, Marx, Klein, and Parrinello \cite{Marx1994AIPIMD,Marx1996AIPIMD,Tuckerman1996PICPMD}.  Early applications to hydrogen bonding in water demonstrated the power of the approach for aqueous systems \cite{Chen2003WaterPIMD}.  Related work on hydrated protons, water interfaces, force-matched PIMD water, and on-the-fly correlated-electronic-structure PIMD further illustrates both the physical insight gained from explicit NQE sampling and the high computational cost of carrying it out directly \cite{Marx1999HydratedProton,Spura2015,Kessler2015Interface,Spura2015CCPIMD}.

A useful compromise is to parametrize flexible, computationally inexpensive water models directly against first-principles forces.  Force matching was originally introduced as a route to derive effective interatomic potentials from reference forces rather than from energies alone \cite{Ercolessi1994}. In earlier work, Spura \emph{et al.} introduced \qtip-like models obtained by force matching to density-functional-theory (DFT) molecular dynamics and used them to study nuclear quantum effects in liquid water by PIMD \cite{Spura2015}.  In a subsequent assessment, K\"oster \emph{et al.} showed that such force-matched models are competitive in reproducing the structure of liquid water and can outperform several common empirical force fields for structural properties, even when their thermodynamic performance is not uniformly empirical-water-model quality \cite{Koester2016}.  A further advantage of this route is specific to PIMD.  Because the reference forces are obtained from classical-nuclei first-principles trajectories, the fitted potential does not implicitly contain nuclear quantum effects through the fitting target.  Ring-polymer sampling can therefore add nuclear quantum fluctuations explicitly, with a reduced risk of artificial double counting.  This is different from many conventional empirical water models, such as TIP4P-type potentials \cite{Abascal2005,Vega2011,Jorgensen1983TIP4P}, whose parameters are adjusted to experimental observables and hence already contain nuclear quantum effects implicitly at the level of the effective classical potential.  Using such models directly in PIMD can therefore mix explicit ring-polymer quantum fluctuations with quantum effects already folded into the empirical parametrization.  Those earlier parametrizations used TPSS-D3 reference data, i.e., a meta-generalized-gradient approximation supplemented by an empirical dispersion correction \cite{Spura2015,Koester2016,Tao2003TPSS,Grimme2010D3}.  Although the TPSS meta-functional uses the kinetic-energy density and therefore goes beyond a conventional generalized-gradient approximation, it does not contain nonlocal Hartree--Fock exact exchange.  These studies established the basic strategy: use accurate first-principles simulations to generate reference configurations and forces, fit a flexible analytic water potential, and then exploit the fitted potential for long classical and path-integral simulations.

An important remaining question is how far the quality of the reference electronic structure can be increased without losing the practical advantage of the force-matched model.  Semilocal density functionals are affordable and have therefore been natural references for early parametrizations, but the structure and dynamics of water are sensitive to the balance of exchange, delocalization error, and hydrogen-bond strength.  The Perdew--Burke--Ernzerhof (PBE) generalized-gradient approximation is widely used for liquid-water benchmarks, whereas the PBE0 hybrid functional includes a fraction of Hartree--Fock exact exchange and therefore provides a more demanding reference for the forces \cite{Perdew1996,Adamo1999}.  Exact exchange partly reduces the self-interaction and excessive delocalization tendencies of semilocal functionals, improves the description of water clusters and liquid-water structure \cite{Todorova2006}, and is therefore an attractive ingredient in a force-matching reference.  This step is computationally much more demanding than moving among semilocal or meta-GGA functionals, especially for large periodic water cells, because the exact-exchange operator is nonlocal and requires substantially more expensive exchange builds.  Consequently, an efficient parametrization strategy is needed not only to improve the numerical conditioning of the force-field fit, but also to minimize the number of costly PBE0 reference calculations that must be generated.  Its direct evaluation in periodic plane-wave or mixed-basis first-principles MD is, however, substantially more expensive than semilocal DFT.  CP2K reduces this bottleneck through its efficient Hartree--Fock exchange infrastructure and related approximations such as the auxiliary density matrix method (ADMM), which can lower the hybrid-functional overhead for condensed-phase simulations \cite{Kuehne2020CP2K,Iannuzzi2026CP2KSimple,Guidon2010ADMM}.  In the present context, ADMM and the projected treatment of linear force-field parameters are complementary: the former reduces the cost of each hybrid-functional force evaluation, whereas the latter reduces the amount of nonlinear search needed to convert those forces into a stable analytic potential.  Once the hybrid-functional forces have been generated, their cost is paid only during the training stage, whereas the fitted potential can be used for long classical and path-integral trajectories.

The present work addresses a technical but important bottleneck in this strategy.  Force-field parametrization contains different types of unknowns.  Some parameters enter the force field linearly, such as amplitudes of fixed basis functions or prefactors once the nonlinear shape parameters have been specified.  Other parameters, such as exponential decay constants, charge-site positions, or repulsive length scales, enter nonlinearly.  Treating all of them as generic nonlinear optimization variables is inefficient and can be numerically fragile because parameters of different physical character and scale compete in the same search.  This is particularly problematic for water, where small changes in short-range O--O repulsion, charge distribution, and intramolecular flexibility can produce visibly different radial distribution functions (RDFs) and nuclear quantum effects.

We therefore recast force matching as a separable nonlinear least-squares problem.  For every trial set of nonlinear parameters, the optimal linear coefficients are obtained from a weighted linear least-squares problem.  Only the projected residual is passed to the nonlinear optimizer.  This idea follows the \varpro{} method of Golub and Pereyra for nonlinear least-squares problems whose variables separate \cite{Golub1973,Golub2003,OLeary2013}.  In the present context, the projection is not merely a numerical convenience.  It reflects the physical structure of an analytic force field: nonlinear parameters define the shape of the interaction functions, whereas linear parameters determine how strongly those shapes contribute to the forces.  Separating these two roles makes it possible to compare different potential forms, in particular Lennard-Jones (LJ) and Buckingham descriptions of the O--O repulsion, without allowing the optimizer to hide deficiencies of one term through unstable compensation by others.

The goal of this paper is thus twofold.  First, we use CP2K-based PBE0 hybrid-functional reference forces that explicitly contain the exact-exchange contribution, thereby moving the reference level beyond the semilocal DFT data used for earlier force-matched water models.  Second, we formulate and document a modified force-matching protocol with explicit separation of linear and nonlinear parameters and apply it to derive new \qtip-like water potentials.  We then assess whether the resulting models retain the structural quality required for efficient classical and path-integral simulations.  This positioning is well suited to molecular simulation: the contribution is a reproducible parametrization workflow that converts expensive electronic-structure forces into inexpensive, simulation-ready potentials, while retaining direct tests against both first-principles and experimental liquid-water structure.

\section{Theory and parametrization strategy}

\subsection{Force matching}

Let $\vect R_s$ denote the Cartesian coordinates of snapshot $s$, and let $\vect F^{\mathrm{ref}}_s$ be the corresponding first-principles reference forces.  For a force field with parameter vector $\bm\theta$, force matching minimizes the weighted residual
\begin{equation}
  \chi^2(\bm\theta)
  =
  \sum_{s=1}^{N_{\mathrm{snap}}}
  \left\|
    \mat W_s
    \left[
      \vect F^{\mathrm{ref}}_s
      -
      \vect F^{\mathrm{model}}(\vect R_s;\bm\theta)
    \right]
  \right\|_2^2 ,
  \label{eq:force_matching}
\end{equation}
where $\mat W_s$ contains optional component weights.  In the simplest case all Cartesian force components are weighted equally.  More generally, weights can be used to balance intermolecular and intramolecular forces, avoid overemphasizing high-force configurations, or regularize components whose reference uncertainty differs.

The model forces are the negative gradient of an analytic potential energy,
\begin{equation}
  \vect F^{\mathrm{model}}(\vect R;\bm\theta)
  =
  -\nabla_{\vect R} U(\vect R;\bm\theta).
  \label{eq:model_forces}
\end{equation}
The force field used here follows the \qtip{} philosophy: the water molecule is flexible, the O--H stretch is represented by an anharmonic intramolecular term, the H--O--H bend is flexible, and electrostatics are described by a four-site charge distribution.  The intermolecular short-range O--O interaction is represented either by an LJ or a Buckingham-type term.  The relevant total energy can be written schematically as
\begin{equation}
\begin{split}
  U &=
  \sum_{i=1}^{N_{\mathrm{mol}}}
  \left[
    \sum_{\lambda=1}^{2}
    U_{\mathrm{str}}(r_{i\lambda})
    +
    U_{\mathrm{bend}}(\theta_i)
  \right]
  \\
  &\quad+
  \sum_{1\le i<j\le N_{\mathrm{mol}}}
  \left[
    U_{\mathrm{OO}}(R_{ij})
    +
    U_{\mathrm{el}}(\vect R_i,\vect R_j)
  \right],
\end{split}
  \label{eq:potential}
\end{equation}
where $r_{i\lambda}$ are the two O--H distances in molecule $i$, $\theta_i$ is its H--O--H angle, $U_{\mathrm{str}}$ denotes the flexible stretching potential, $U_{\mathrm{bend}}$ the angular term, $U_{\mathrm{OO}}$ the oxygen--oxygen short-range interaction, and $U_{\mathrm{el}}$ the intermolecular Coulomb interaction including the massless charge site of the four-site model.
The intramolecular flexibility is represented by the same functional philosophy as \qtip{}, namely a Morse-like O--H stretch \cite{Habershon2009}:
The Morse form is intrinsically anharmonic, which is important for water because the O--H stretching coordinate responds asymmetrically to zero-point motion, isotope substitution, and hydrogen-bond-induced bond elongation.  This anharmonic flexibility was a central ingredient of \qtip{} and of subsequent force-matched PIMD water models \cite{Spura2015,Koester2016,Habershon2009}.
\begin{equation}
  U_{\mathrm{str}}(r)
  =
  D_{\mathrm{OH}}
  \left[
    1-\exp[-a_{\mathrm{OH}}(r-r_{\mathrm{OH}}^0)]
  \right]^2 ,
  \label{eq:morse}
\end{equation}
and a harmonic or weakly anharmonic angular restoring term,
\begin{equation}
  U_{\mathrm{bend}}(\theta)
  =
  \frac{1}{2}
  k_\theta
  \left(\theta-\theta_0\right)^2 .
  \label{eq:bend}
\end{equation}
In the production implementation used below, the Morse-like O--H stretch is evaluated in the same quartic-expanded form as in the earlier force-matched water potentials and in \texttt{RPMD\_Mainz},
\begin{equation}
  U_{\mathrm{str}}^{(4)}(r)
  =
  D_r
  \left[
    \alpha_r^2 \Delta r^2
    -
    \alpha_r^3 \Delta r^3
    +
    \frac{7}{12}\alpha_r^4 \Delta r^4
  \right],
  \qquad
  \Delta r=r-r_{\mathrm{OH}}^0 ,
  \label{eq:morse_quartic}
\end{equation}
and the bending contribution is
\begin{equation}
  U_{\mathrm{bend}}(\theta)=b_\theta(\theta-\theta_0)^2 ,
  \label{eq:bend_rpmd}
\end{equation}
where $b_\theta$ corresponds to one half of the conventional harmonic force constant.
The intermolecular electrostatics are evaluated between fixed partial charges located on the hydrogen atoms and on a massless site displaced from the oxygen atom along the instantaneous molecular bisector.  For molecule $i$ with oxygen position $\vect r_{O_i}$ and hydrogen positions $\vect r_{H_{i1}}$ and $\vect r_{H_{i2}}$, the fourth site is written as
\begin{equation}
  \vect r_{M_i}
  =
  \gamma \vect r_{O_i}
  +
  \frac{1-\gamma}{2}
  \left(
    \vect r_{H_{i1}}+\vect r_{H_{i2}}
  \right),
  \label{eq:msite}
\end{equation}
where $\gamma$ controls the displacement of the negative charge site from the oxygen toward the hydrogen bisector.  The site carries charge $q_M=-2q_H$, while each hydrogen carries $q_H$, so that the molecule remains neutral.  In the parameter-file notation used in Table~\ref{tab:pbe0_oogam_parameters}, the negative charge is denoted $q_o$ and the hydrogen charges are $q_H=-q_o/2$; the M-site position is obtained from Eq.~\eqref{eq:msite} with $\gamma=\texttt{alpha}$.  The electrostatic part is then
\begin{equation}
  U_{\mathrm{el}}
  =
  \sum_{i<j}
  \sum_{a\in\{H_{i1},H_{i2},M_i\}}
  \sum_{b\in\{H_{j1},H_{j2},M_j\}}
  \frac{q_a q_b}{4\pi\varepsilon_0 r_{ab}},
  \label{eq:electrostatics}
\end{equation}
with the usual Ewald treatment under periodic boundary conditions in condensed-phase simulations.  The charge-site parameter $\gamma$ is nonlinear because it changes both the magnitude and direction of the electrostatic forces generated by a given nuclear geometry.  By contrast, several prefactors in the stretching, bending, and short-range intermolecular terms can be treated as linear once the nonlinear shape parameters have been specified.

\subsection{Path-integral sampling and ring-polymer contraction}

The fitted model is intended for quantum-nuclear simulations.  In PIMD, the quantum canonical distribution is sampled by replacing each nucleus $i$ by a cyclic polymer of $P$ beads.  The corresponding classical ring-polymer Hamiltonian is
\begin{equation}
  H_P =
  \sum_{p=1}^{P}
  \sum_i
  \left[
    \frac{\left(\vect p_i^{(p)}\right)^2}{2m_i}
    +
    \frac{1}{2}
    m_i \omega_P^2
    \left|
      \vect r_i^{(p)}-\vect r_i^{(p+1)}
    \right|^2
  \right]
  +
  \frac{1}{P}
  \sum_{p=1}^{P}
  U\!\left(\vect R^{(p)}\right),
  \label{eq:ring_polymer_hamiltonian}
\end{equation}
where $\omega_P=P/(\beta\hbar)$ and $\vect r_i^{(P+1)}=\vect r_i^{(1)}$.  The classical limit is recovered for $P=1$, whereas the quantum simulations reported below use a finite-bead discretization of the nuclear path.

For a \qtip-like potential, the force hierarchy is particularly well suited to ring-polymer contraction (RPC) \cite{Habershon2009,Markland2008RPC,Markland2008Electrostatics}.  The intramolecular Morse and bending terms vary rapidly along the imaginary-time path and are therefore evaluated on the full ring polymer.  Slower intermolecular contributions, especially long-ranged electrostatics, can be evaluated on a contracted ring polymer with $P'<P$ beads obtained by retaining only the lowest normal modes of the full ring polymer.  If
\begin{equation}
  U =
  U_{\mathrm{fast}} + U_{\mathrm{slow}},
  \label{eq:rpc_split}
\end{equation}
then the potential part of the ring-polymer Hamiltonian is approximated as
\begin{equation}
  \frac{1}{P}
  \sum_{p=1}^{P}
  U\!\left(\vect R^{(p)}\right)
  \approx
  \frac{1}{P}
  \sum_{p=1}^{P}
  U_{\mathrm{fast}}\!\left(\vect R^{(p)}\right)
  +
  \frac{1}{P'}
  \sum_{p'=1}^{P'}
  U_{\mathrm{slow}}\!\left(\widetilde{\vect R}^{(p')}\right),
  \label{eq:rpc}
\end{equation}
where $\widetilde{\vect R}^{(p')}$ denotes the contracted bead coordinates.  This is not required to define the force-matched potential, but it explains why a flexible analytic model can turn expensive reference information into long PIMD trajectories: the high-frequency intramolecular quantum fluctuations are retained explicitly, while slowly varying intermolecular forces can be evaluated on fewer beads when RPC is used.

\subsection{Separation of linear parameters}

The central observation is that many analytic force-field terms can be expressed as a linear combination of basis contributions once the nonlinear shape parameters are fixed.  We write
\begin{equation}
  U(\vect R;\bm\alpha,\vect c)
  =
  \sum_{k=1}^{n_{\mathrm{lin}}}
  c_k \, \Phi_k(\vect R;\bm\alpha)
  +
  U_0(\vect R;\bm\alpha),
  \label{eq:separable_energy}
\end{equation}
where $\bm\alpha$ contains nonlinear parameters and $\vect c$ the linear coefficients.  Taking derivatives gives
\begin{equation}
  \vect F^{\mathrm{model}}(\vect R;\bm\alpha,\vect c)
  =
  \sum_{k=1}^{n_{\mathrm{lin}}}
  c_k \, \vect f_k(\vect R;\bm\alpha)
  +
  \vect f_0(\vect R;\bm\alpha),
  \label{eq:separable_force}
\end{equation}
with $\vect f_k=-\nabla_{\vect R}\Phi_k$ and $\vect f_0=-\nabla_{\vect R}U_0$.

After collecting all weighted force components into a single vector $\vect y$ and the corresponding model force basis into a matrix $\mat A(\bm\alpha)$, Eq.~\eqref{eq:force_matching} becomes
\begin{equation}
  \chi^2(\bm\alpha,\vect c)
  =
  \left\|
    \vect y
    -
    \mat A(\bm\alpha)\vect c
    -
    \vect b(\bm\alpha)
  \right\|_2^2 .
  \label{eq:linearized_residual}
\end{equation}
For any fixed $\bm\alpha$, the optimal linear parameters satisfy
\begin{equation}
  \vect c^\star(\bm\alpha)
  =
  \arg\min_{\vect c}
  \left\|
    \vect y
    -
    \vect b(\bm\alpha)
    -
    \mat A(\bm\alpha)\vect c
  \right\|_2^2 .
  \label{eq:linear_problem}
\end{equation}
Using the Moore--Penrose pseudoinverse or a rank-revealing QR or singular-value-decomposition (SVD) solver, this gives
\begin{equation}
  \vect c^\star(\bm\alpha)
  =
  \mat A^+(\bm\alpha)
  \left[
    \vect y-\vect b(\bm\alpha)
  \right].
  \label{eq:pseudoinverse}
\end{equation}
The nonlinear optimizer then minimizes the projected objective
\begin{equation}
  \tilde{\chi}^2(\bm\alpha)
  =
  \left\|
    \left[
      \mat I-\mat A(\bm\alpha)\mat A^+(\bm\alpha)
    \right]
    \left[
      \vect y-\vect b(\bm\alpha)
    \right]
  \right\|_2^2 .
  \label{eq:projected_residual}
\end{equation}
This is the force-field analogue of the classical variable-projection reduction.  It has three practical advantages.  First, the nonlinear optimizer explores a lower-dimensional parameter space.  Second, the best linear response to a trial nonlinear model is always used, avoiding unnecessary coupling between amplitudes and shape parameters.  Third, ill-conditioned linear combinations are exposed directly by the singular spectrum of $\mat A$, rather than being hidden inside a nonlinear optimization trajectory.

\subsection{Projected fitting workflow}

The complete fitting procedure is summarized in Table~\ref{tab:workflow}.  In practice, the most important implementation detail is that the linear least-squares problem is solved repeatedly and deterministically inside the nonlinear search.  The nonlinear optimizer therefore never sees a parameter vector in which, for example, a short-range amplitude has not yet relaxed to the current decay length.  This is the main practical difference from a simultaneous nonlinear fit of all parameters.

\begin{table}[t]
\centering
\caption{Projected force-matching workflow used in the present parametrization.}
\label{tab:workflow}
\begin{tabular}{p{0.16\linewidth}p{0.74\linewidth}}
\toprule
Step & Operation \\
\midrule
1 & Generate a first-principles reference trajectory and collect Cartesian forces for representative liquid-water snapshots. \\
2 & Choose a \qtip-like analytic potential form and partition its parameters into nonlinear shape variables $\bm\alpha$ and linear amplitudes $\vect c$. \\
3 & For a trial $\bm\alpha$, assemble the weighted force-basis matrix $\mat A(\bm\alpha)$ and residual offset $\vect b(\bm\alpha)$. \\
4 & Solve the linear least-squares problem for $\vect c^\star(\bm\alpha)$ using a rank-aware QR or SVD solver. \\
5 & Return the projected residual $\tilde{\chi}^2(\bm\alpha)$ to the nonlinear optimizer. \\
6 & Validate the resulting model by classical and path-integral simulations, with site--site radial distribution functions as the primary structural metric. \\
\bottomrule
\end{tabular}
\end{table}

The projection also provides a useful diagnostic for over-parametrization.  If several basis functions are nearly linearly dependent over the training set, the singular values of $\mat A$ reveal the redundancy directly.  Such terms can then be removed or regularized before production fitting.  This diagnostic is particularly useful for flexible water models because intramolecular and intermolecular force components are not always cleanly separated in instantaneous liquid configurations.

\subsection{LJ and Buckingham oxygen--oxygen terms}

The short-range O--O interaction is especially important in liquid water because it controls the first peak position and width of the O--O RDF and therefore the local tetrahedral network.  In many empirical models this term is represented by an LJ potential,
\begin{equation}
  U_{\mathrm{LJ}}(R)
  =
  4\varepsilon
  \left[
    \left(\frac{\sigma}{R}\right)^{12}
    -
    \left(\frac{\sigma}{R}\right)^6
  \right].
  \label{eq:lj}
\end{equation}
The repulsive $R^{-12}$ term is computationally convenient but not derived from the exponential character of short-range electronic repulsion.  We therefore also consider the Buckingham form
\begin{equation}
  U_{\mathrm{Buck}}(R)
  =
  A\exp(-B R)
  -
  \frac{C}{R^6},
  \label{eq:buckingham}
\end{equation}
where $A$ and $C$ are linear once the exponential decay parameter $B$ is fixed.  This makes Eq.~\eqref{eq:buckingham} naturally suited to the separable optimization described above.  The nonlinear parameter controls the range of the repulsion, while the optimal amplitude is obtained analytically at each nonlinear step.
For the final PBE0-oogam parametrization, the Buckingham term was used in the reparametrized form implemented in \texttt{RPMD\_Mainz},
\begin{equation}
  U_{\mathrm{oogam}}(R)
  =
  A_{\mathrm{OO}}\exp\!\left(-\gamma_{\mathrm{OO}}\frac{R}{\sigma_{\mathrm{OO}}}\right)
  -
  \frac{C_{6,\mathrm{OO}}}{R^6},
  \label{eq:oogam}
\end{equation}
with
\begin{equation}
  A_{\mathrm{OO}}
  =
  \frac{6\varepsilon_{\mathrm{OO}}\exp(\gamma_{\mathrm{OO}})}
       {\gamma_{\mathrm{OO}}-6},
  \qquad
  C_{6,\mathrm{OO}}
  =
  \frac{\varepsilon_{\mathrm{OO}}\sigma_{\mathrm{OO}}^6}
       {1-6/\gamma_{\mathrm{OO}}}.
  \label{eq:oogam_coefficients}
\end{equation}
This parametrization keeps $\sigma_{\mathrm{OO}}$ and $\varepsilon_{\mathrm{OO}}$ directly interpretable as the O--O minimum position and well depth of the short-range term, while $\gamma_{\mathrm{OO}}$ controls the steepness of the exponential wall.

\section{Computational details}

The reference data and simulation protocol follow the earlier CP2K-based force-matching studies of liquid water where appropriate \cite{Spura2015,Koester2016}.  Reference configurations were sampled from first-principles MD of liquid water at ambient conditions.  The simulations used periodic boundary conditions and a cubic simulation cell containing 216 water molecules at the experimental density.  Electronic-structure calculations were performed with CP2K, using the Gaussian and plane-wave framework of the \textsc{Quickstep} module \cite{Kuehne2020CP2K}.  The Perdew--Burke--Ernzerhof (PBE) functional 
\cite{Perdew1996} was used for the validation against the previous parametrization protocol, and the PBE0 hybrid functional 
\cite{Adamo1999}, including the exact-exchange component, was used for the improved production parametrization.  Thus, the final fitted potential is trained not merely on a semilocal DFT reference but on a hybrid-functional force field that contains explicit Hartree--Fock exchange.  The present parametrization otherwise follows the computational details of the earlier Molecular Physics study for the reference trajectory generation and force evaluation, including the simulation cell, sampling protocol, and remaining electronic-structure settings, so that changes in the fitted model can be attributed to the improved reference level and modified optimization strategy rather than to unrelated changes in the training protocol.

In CP2K, the practical use of exact exchange for condensed-phase systems benefits from localized Gaussian orbitals, integral screening, efficient Poisson solvers, and ADMM \cite{Kuehne2020CP2K,Iannuzzi2026CP2KSimple,Guidon2010ADMM}.  In ADMM, the expensive Hartree--Fock exchange energy of the full density matrix $P$ is approximated by evaluating exact exchange for a projected auxiliary density matrix $\hat P$ in a smaller basis and correcting the difference with a semilocal exchange functional,
\begin{equation}
  E_x^{\mathrm{HFX}}[P]
  \approx
  E_x^{\mathrm{HFX}}[\hat P]
  +
  E_x^{\mathrm{DFT}}[P]
  -
  E_x^{\mathrm{DFT}}[\hat P] .
  \label{eq:admm}
\end{equation}
This approximation retains the main physical benefit of including explicit exchange while avoiding the full cost of evaluating four-center exchange integrals in the production basis at every molecular-dynamics step.  The use of hybrid-functional reference forces is therefore feasible for generating a finite force-matching data set, while the resulting analytic water potential removes the hybrid-functional cost completely from subsequent classical and path-integral sampling.

The training set contained 1500 snapshots.  For each snapshot, all Cartesian force components were included in the least-squares objective.  The force field was fitted in stages.  First, the separable optimization was tested against PBE reference data, for which previously fitted models and direct first-principles benchmarks are available.  This validation establishes that the modified optimizer reproduces the earlier force-matched structural reference while improving numerical stability.  Second, the same protocol was applied to PBE0 hybrid-functional reference forces, with particular attention to the O--O short-range term and to the sensitivity of the liquid structure to LJ versus Buckingham repulsion.  This yielded both the original PBE0-LJ fit and the PBE0-oogam fit; the latter differs primarily in the O--O short-range term, while leaving the intramolecular parameters and charge-site geometry nearly unchanged.

Classical MD and PIMD simulations were then carried out with the optimized potentials.  The production trajectories reported here used the \texttt{RPMD\_Mainz} code developed for flexible water models, ring-polymer molecular dynamics, and ring-polymer contraction simulations \cite{RPMDMainz}.  The PIMD simulations used the standard ring-polymer representation of the quantum canonical partition function \cite{Feynman1965,Tuckerman1993,Ceriotti2010}, with a flexible intramolecular water model so that zero-point motion in the O--H stretches and H--O--H bend is retained.  Classical reference simulations correspond to $P=1$, whereas quantum-nuclear simulations used the same finite-bead setup as the earlier \qtip{} studies.  When ring-polymer contraction was used, the bonded terms were evaluated on the full ring polymer, while the smoother intermolecular terms were evaluated on contracted rings following the hierarchy in Eq.~\eqref{eq:rpc}.  The earlier TPSS-D3 force-matched model, the PBE validation model, the original PBE0-LJ fit, and the final PBE0-oogam parameter set are listed in Table~\ref{tab:pbe0_oogam_parameters}.  The TPSS-D3 and PBE parameters are taken from the earlier force-matching work and its \texttt{RPMD\_Mainz} parameter files \cite{Spura2015,Koester2016,RPMDMainz}.  The notation follows the \texttt{RPMD\_Mainz} parameter-file convention: all lengths are in bohr, energies in Hartree, charges in units of the elementary charge, masses in electron masses, and the equilibrium angle is supplied in degrees and converted internally to radians.

\begin{table}[t]
\centering
\scriptsize
\caption{Force-matched water-model parameter sets used for comparison and production simulations.  Keyword names follow the \texttt{RPMD\_Mainz} input format.}
\label{tab:pbe0_oogam_parameters}
\resizebox{\linewidth}{!}{%
\begin{tabular}{lllllll}
\toprule
Keyword & Meaning & TPSS-D3 & PBE & PBE0-LJ & PBE0-oogam & Units / note \\
\midrule
\texttt{wmass} & Molecular mass & 32831.2525000000 & 32831.2525000000 & 32831.2525000000 & 32831.2525000000 & $m_e$ \\
\texttt{omass} & Oxygen mass & 29156.9471000000 & 29156.9471000000 & 29156.9471000000 & 29156.9471000000 & $m_e$ \\
\texttt{hmass} & Hydrogen mass & 1837.1527000000 & 1837.1527000000 & 1837.1527000000 & 1837.1527000000 & $m_e$ \\
\texttt{qo} & Negative M-site charge $q_o$ & -1.03181231055394 & -1.13471198989530 & -1.09045545732501 & -1.09058833465870 & $e$ \\
\texttt{alpha} & M-site displacement $\gamma$ & 0.729814982301420 & 0.655510428642656 & 0.706724657372371 & 0.707292119168911 & dimensionless \\
\texttt{oo\_sig} & O--O minimum parameter $\sigma_{\mathrm{OO}}$ & 5.97820609490816 & 5.97460289857713 & 6.79682208332826 & 7.07528725322939 & bohr \\
\texttt{oo\_eps} & O--O well depth $\varepsilon_{\mathrm{OO}}$ & 0.0002513613860605 & 0.0002268088165733 & 0.0000410349712031 & 0.0000628218329731 & Hartree \\
\texttt{oo\_gam} & Buckingham steepness $\gamma_{\mathrm{OO}}$ & 0.0000000000000000 & 0.0000000000000000 & 0.0000000000000000 & 17.9071323631243 & dimensionless \\
\texttt{thetad} & Equilibrium angle $\theta_0$ & 107.378313652498 & 107.406234838648 & 108.407496717120 & 108.413682151614 & degree \\
\texttt{reoh} & Equilibrium O--H distance $r_{\mathrm{OH}}^0$ & 1.82733537724340 & 1.82767168120489 & 1.80273088698414 & 1.80270180269420 & bohr \\
\texttt{apot} & Stretch parameter $D_r$ & 0.161988809510218 & 0.162486065049365 & 0.155227124325556 & 0.155216143017800 & Hartree \\
\texttt{bpot} & Bend parameter $b_\theta$ & 0.0634051341058466 & 0.0617059503139460 & 0.0643095038433985 & 0.0643298280811618 & Hartree rad$^{-2}$ \\
\texttt{alp} & Stretch inverse length $\alpha_r$ & 1.23158483101631 & 1.23241936757886 & 1.32935428569046 & 1.32963513669870 & bohr$^{-1}$ \\
\bottomrule
\end{tabular}
}
\end{table}

Structural properties were analyzed through the site--site RDFs
\begin{equation}
  g_{\alpha\beta}(r)
  =
  \frac{1}{4\pi r^2 \rho_\beta N_\alpha}
  \left\langle
    \sum_{i=1}^{N_\alpha}
    \sum_{j=1}^{N_\beta}{}^{\prime}
    \delta(r-r_{ij})
  \right\rangle ,
  \label{eq:rdf}
\end{equation}
where the prime excludes self terms when $\alpha=\beta$.  For orthorhombic simulation cells, efficient algorithms for evaluating such pair-correlation functions were discussed by R\"ohrig and K\"uhne \cite{Rohrig2013}.  The computed O--O, O--H, and H--H RDFs were compared with the first-principles reference and with neutron-scattering-based experimental distributions \cite{Soper2013}.

\section{Results and discussion}

\subsection{Effect of the projected optimization}

\begin{figure*}[t]
  \centering
  \begin{subfigure}{0.32\textwidth}
    \centering
    \includegraphics[width=\linewidth]{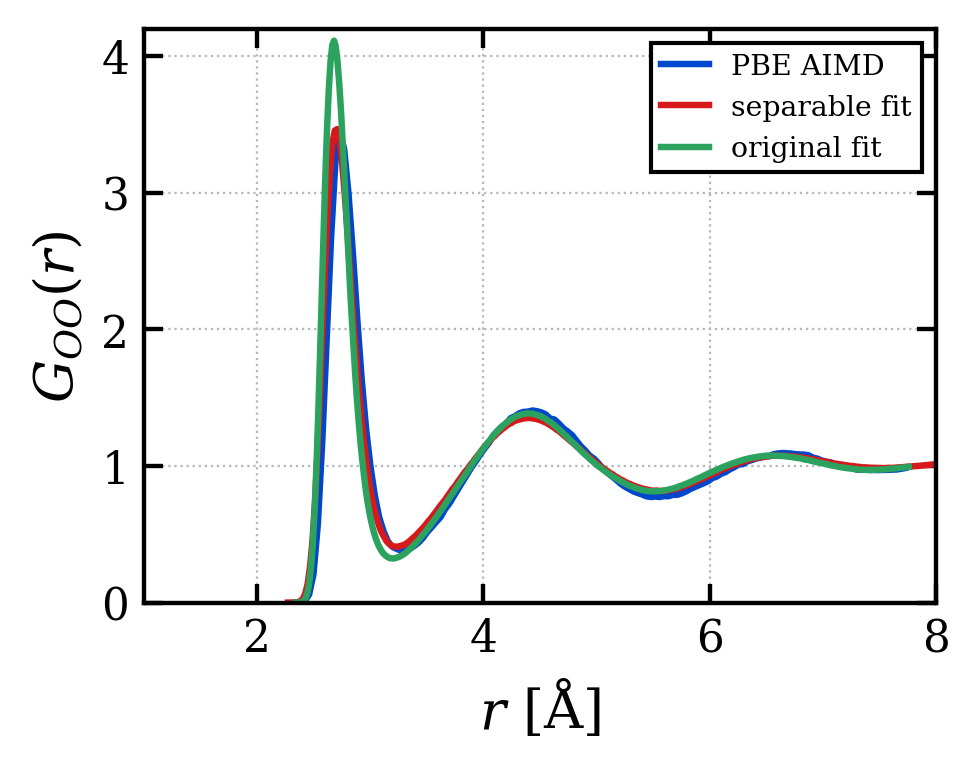}
    \caption{O--O}
  \end{subfigure}
  \begin{subfigure}{0.32\textwidth}
    \centering
    \includegraphics[width=\linewidth]{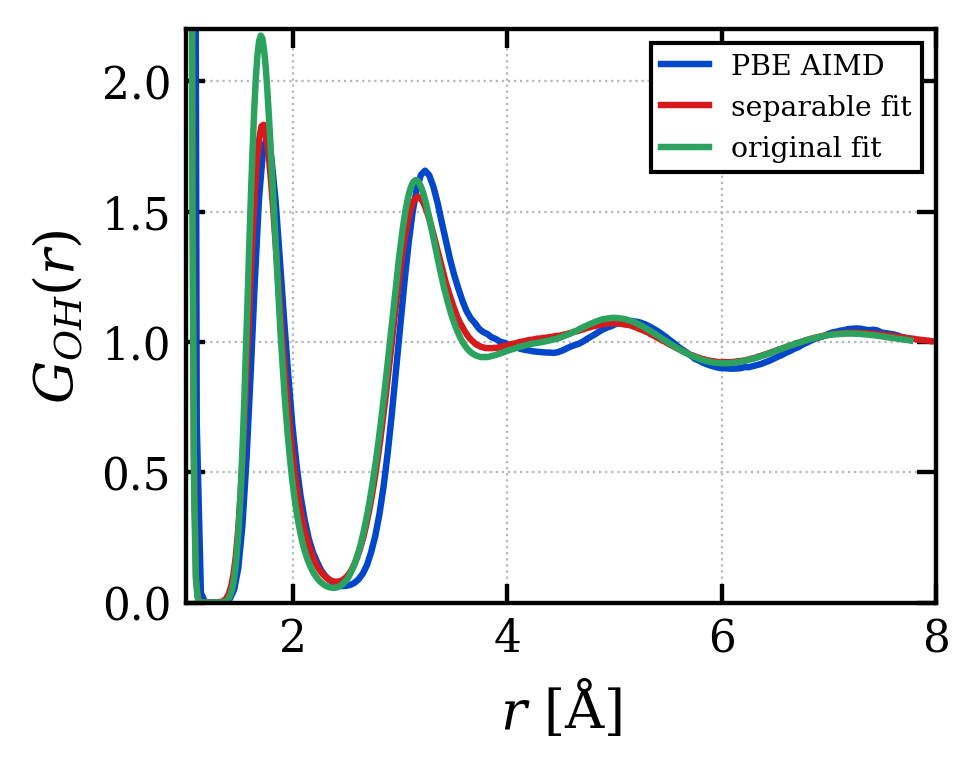}
    \caption{O--H}
  \end{subfigure}
  \begin{subfigure}{0.32\textwidth}
    \centering
    \includegraphics[width=\linewidth]{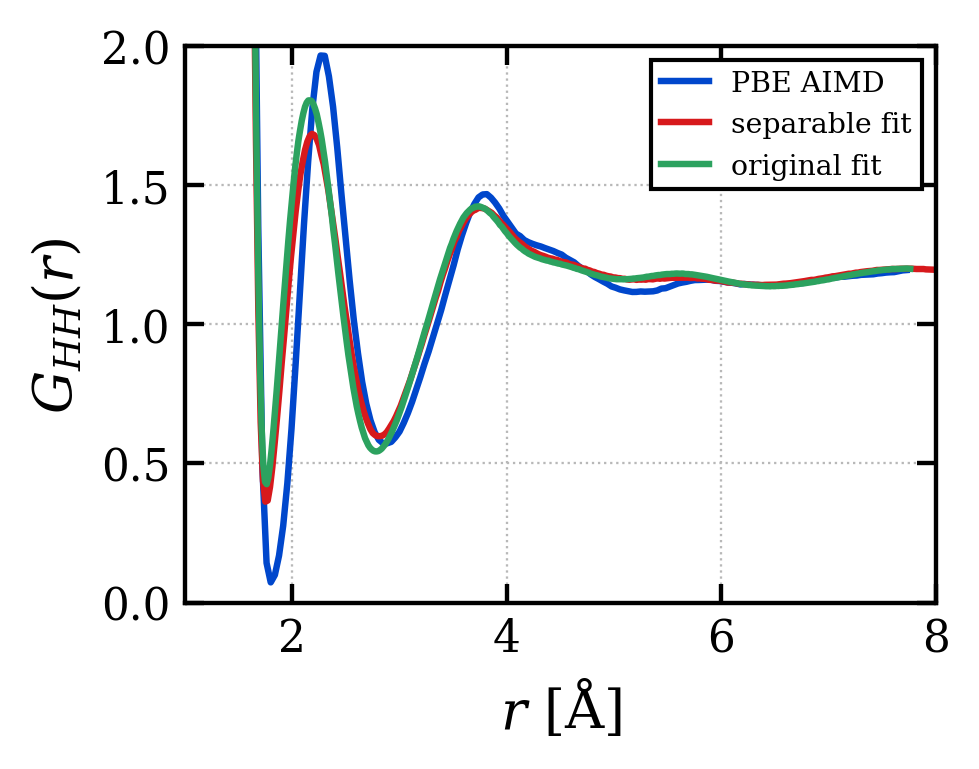}
    \caption{H--H}
  \end{subfigure}
  \caption{Benchmark of the force-matched water model against the PBE-based reference using the original and modified optimization strategies.  The projected optimization eliminates the linear coefficients at each nonlinear step and thereby separates the choice of functional form from the determination of the optimal amplitudes.}
  \label{fig:projection_benchmark}
\end{figure*}

Figure~\ref{fig:projection_benchmark} compares the structural performance of the earlier and modified force-matching procedures for the PBE reference.  The purpose of this test is not to introduce a new electronic-structure reference, but to isolate the effect of the optimization.  The projected treatment of the linear parameters yields a fit that tracks the reference structure without requiring manual compensation among strongly coupled nonlinear and linear terms.  In practice, the reduction of the nonlinear parameter space makes the optimization less sensitive to the starting guess and makes it easier to diagnose whether a poor result originates from the potential-energy form or from a local optimization minimum.

This distinction matters because force matching is often underdetermined in a practical sense: different combinations of short-range, electrostatic, and intramolecular parameters can yield similar force errors over the finite training set but noticeably different liquid structures.  A robust parametrization protocol should therefore not only reduce the force residual, but also avoid artificial cancellation between physically distinct terms.  The separable formulation helps in this respect by making the optimal linear response explicit and by turning the remaining optimization into a search over physically interpretable shape parameters.

\subsection{Structural validation against first-principles and experimental references}

\begin{figure*}[t]
  \centering
  \includegraphics[width=0.96\textwidth]{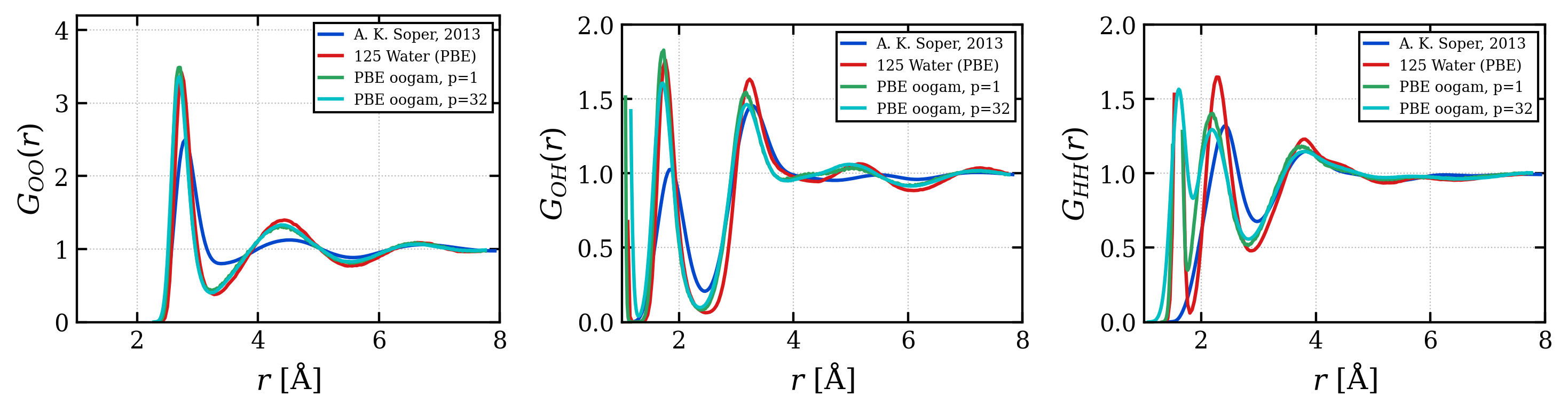}
  \par\vspace{0.3em}
  \makebox[0.32\textwidth]{(a) O--O}%
  \makebox[0.32\textwidth]{(b) O--H}%
  \makebox[0.32\textwidth]{(c) H--H}
  \caption{Site--site radial distribution functions for the PBE validation case.  The comparison tests whether the modified force-matching procedure preserves the structural quality of the first-principles reference while retaining the efficiency of a flexible analytic water potential.}
  \label{fig:pbe_rdfs}
\end{figure*}

The PBE validation is shown in more detail in Fig.~\ref{fig:pbe_rdfs}, where the O--O, O--H, and H--H RDFs are compared with the first-principles reference and the experimental distributions.  The agreement is best judged not by any single peak height, but by whether the model captures the positions and relative widths of the first and second coordination shells.  The O--O distribution is most sensitive to the intermolecular repulsion and hydrogen-bond network topology, whereas the O--H and H--H distributions also reflect intramolecular flexibility and quantum delocalization.

The fitted model reproduces the main features of all three site--site correlations.  In particular, the first O--O shell remains close to the first-principles reference, indicating that the modified optimization does not degrade the hydrogen-bond network.  The O--H and H--H distributions show that the intramolecular flexibility remains compatible with the reference data and with the broadening expected when quantum nuclear fluctuations are included.  These observations are consistent with the earlier conclusion that force-matched flexible water models can reproduce the structure of liquid water more accurately than is often assumed for computationally inexpensive empirical potentials \cite{Koester2016}.

\subsection{Choice of oxygen--oxygen repulsion}

\begin{figure}[t]
  \centering
  \includegraphics[width=0.62\linewidth]{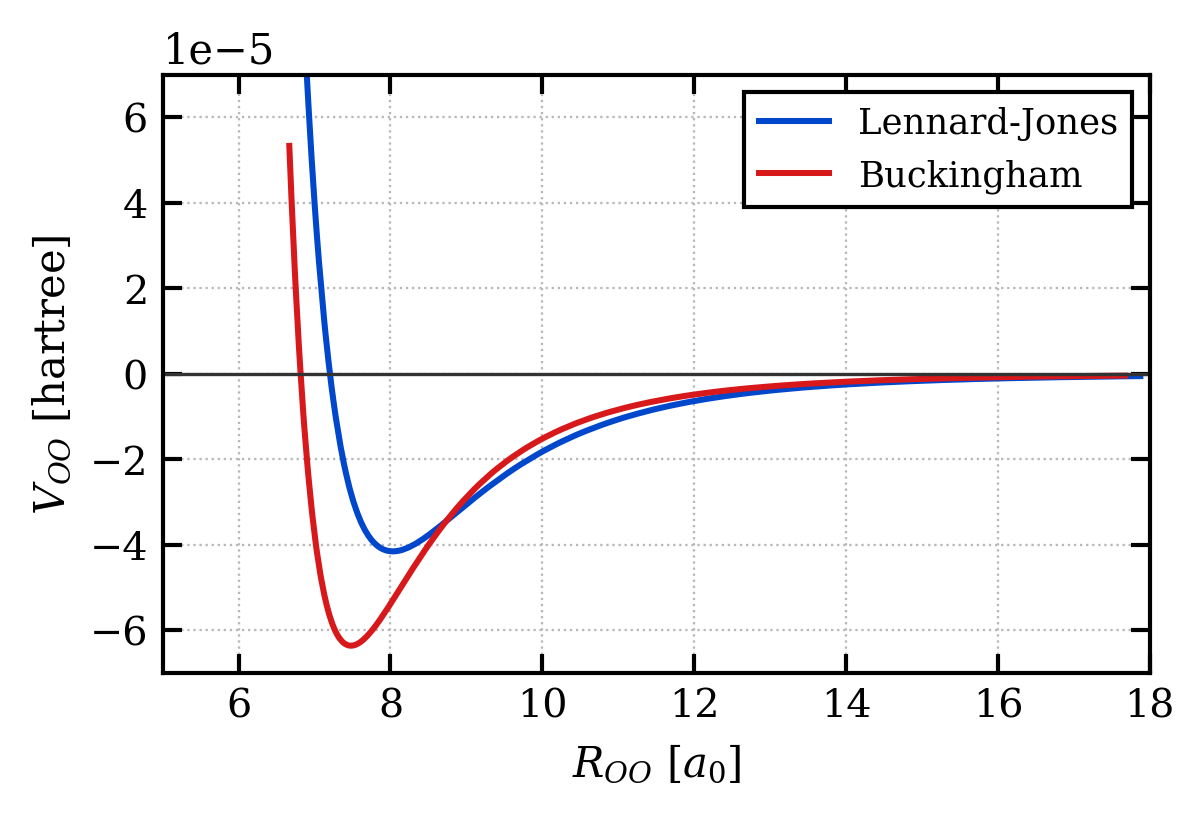}
  \caption{Comparison of Lennard-Jones and Buckingham descriptions of the O--O short-range interaction for the hybrid-functional parametrization.  The Buckingham form separates the exponential repulsive wall from the dispersion-like attractive contribution and is therefore more naturally compatible with the projected linear-parameter optimization.}
  \label{fig:buckingham}
\end{figure}

The comparison between LJ and Buckingham oxygen--oxygen interactions is summarized in Fig.~\ref{fig:buckingham} for the PBE0 hybrid-functional parametrization.  This is the more stringent case because the short-range repulsion and hydrogen-bond forces are inherited from a hybrid-functional reference rather than from a semilocal one.  The Buckingham term provides a more flexible description of the steep short-range repulsion because its exponential wall is controlled independently from the attractive $R^{-6}$ contribution.  In the separable optimization, this distinction is useful because the amplitudes of the repulsive and attractive terms can be optimized linearly for every trial decay length.  The LJ form, by contrast, ties the repulsive and attractive parts more tightly through $\varepsilon$ and $\sigma$.

For water, this additional freedom is not cosmetic.  Small changes in the repulsive wall can shift the first O--O peak, alter the population of interstitial configurations, and modify the balance between intra- and intermolecular nuclear quantum effects.  The fitted Buckingham interaction therefore provides a physically motivated route to improve the short-range structure without abandoning the simplicity of an analytic water model.

\subsection{Best-fit structural models}

\begin{figure*}[t]
  \centering
  \begin{subfigure}{0.32\textwidth}
    \centering
    \includegraphics[width=\linewidth]{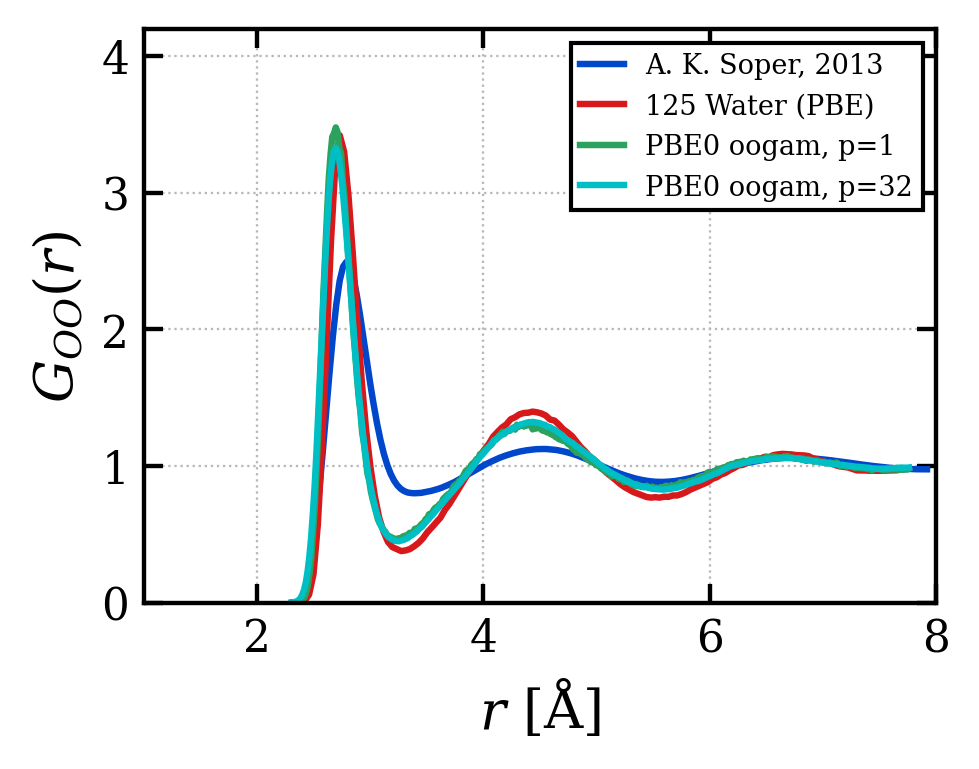}
    \caption{O--O}
  \end{subfigure}
  \begin{subfigure}{0.32\textwidth}
    \centering
    \includegraphics[width=\linewidth]{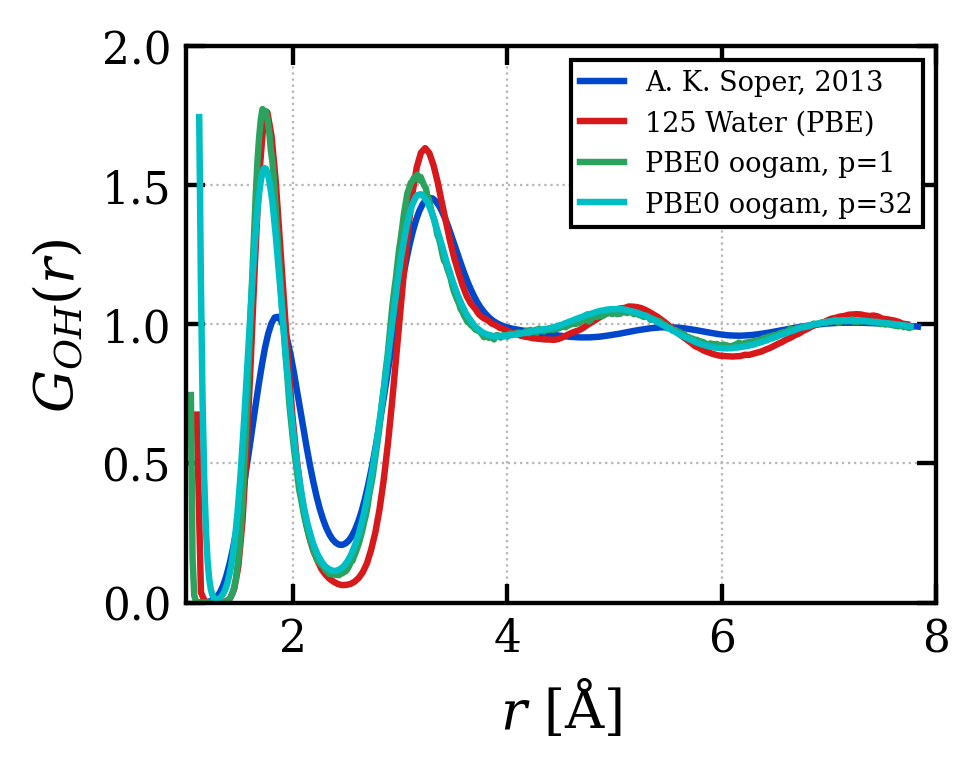}
    \caption{O--H}
  \end{subfigure}
  \begin{subfigure}{0.32\textwidth}
    \centering
    \includegraphics[width=\linewidth]{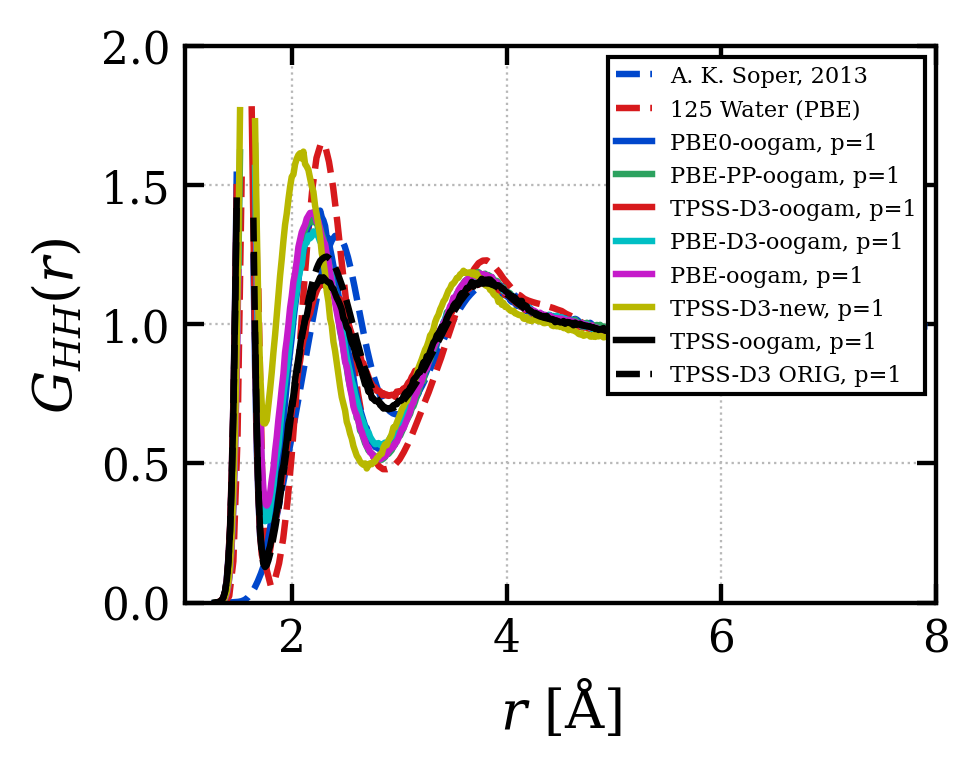}
    \caption{H--H}
  \end{subfigure}
  \caption{Radial distribution functions of the final force-matched water potential.  The figure illustrates the structural quality retained after separating the linear and nonlinear parameters in the optimization.}
  \label{fig:best_rdfs}
\end{figure*}

Figure~\ref{fig:best_rdfs} shows the site--site RDFs obtained from the final fitted potential.  The overall agreement with the reference distributions demonstrates that the modified optimizer can be used not only as a diagnostic tool but as a production parametrization protocol.  The O--O distribution captures the principal coordination-shell structure, while the O--H and H--H distributions remain consistent with a flexible water model suitable for quantum-nuclear sampling.

The remaining deviations should be interpreted in light of the design goal.  The model is not a fully empirical potential tuned independently to density, vaporization enthalpy, dielectric constant, diffusion coefficient, and neutron data.  Instead, it is an electronic-structure-derived model intended to reproduce the local first-principles force field while being inexpensive enough for long classical and path-integral simulations.  The relevant question is therefore whether the analytic model preserves the first-principles structural response sufficiently well to make quantum-nuclear sampling affordable.  On that criterion, the projected force-matching procedure performs well.

\subsection{Path-integral molecular dynamics with the fitted potential}

The final test is whether the force-matched model remains stable and physically meaningful in PIMD.  Flexible water models are essential here because constraining the O--H stretch would remove one of the central channels through which nuclear quantum effects modify the hydrogen-bond network.  Earlier PIMD studies with \qtip{} and force-matched variants showed that quantum fluctuations in water are competing: intramolecular zero-point motion tends to elongate and soften O--H bonds, while intermolecular quantum delocalization can either weaken or strengthen hydrogen bonds depending on the local environment \cite{Spura2015,Habershon2009}.

\begin{figure}[t]
  \centering
  \begin{subfigure}{0.48\linewidth}
    \centering
    \includegraphics[width=\linewidth]{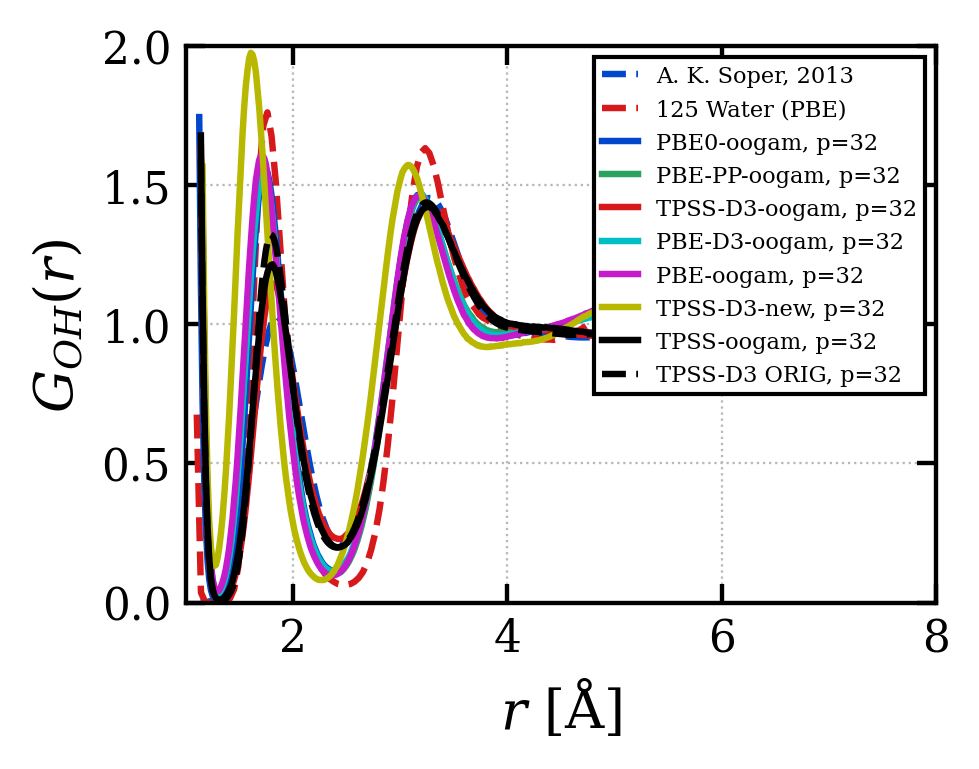}
    \caption{O--H}
  \end{subfigure}
  \begin{subfigure}{0.48\linewidth}
    \centering
    \includegraphics[width=\linewidth]{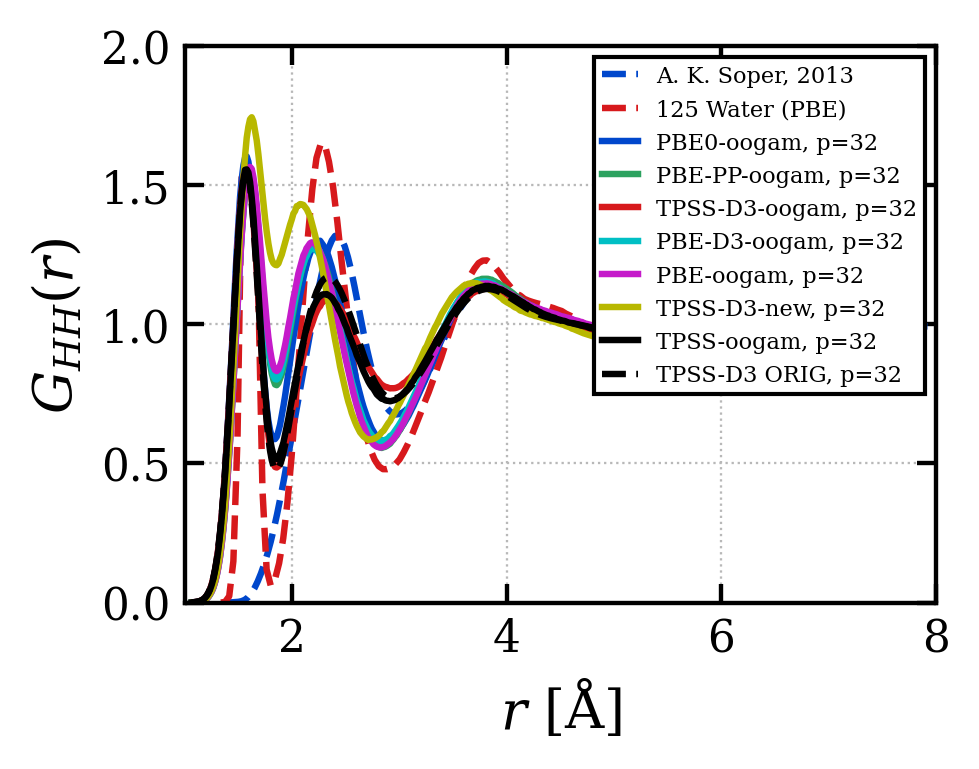}
    \caption{H--H}
  \end{subfigure}
  \caption{Path-integral RDFs for the final fitted potential.  The flexible model remains stable under quantum-nuclear sampling and retains the intramolecular broadening required for \qtip-like simulations.}
  \label{fig:qm_rdfs}
\end{figure}

The present models are designed to retain this balance.  Figure~\ref{fig:qm_rdfs} reports quantum-nuclear RDFs from the final potential.  The quantum simulations preserve the main liquid structure while broadening the distributions in a manner consistent with flexible PIMD sampling.  This behavior is crucial because a force-matched model that reproduces classical RDFs but fails under ring-polymer sampling would not be useful for the intended application.

Taken together, these results support the use of separable force matching as a practical refinement of the original ab initio force-matching strategy.  The gain is methodological rather than merely numerical: by isolating the linear amplitudes from the nonlinear shape parameters, the fit becomes easier to reproduce, easier to diagnose, and better suited to comparing physically distinct potential forms.

\section{Conclusions}

We have reformulated the parametrization of first-principles-based flexible water potentials as a separable nonlinear least-squares problem and combined this optimization strategy with PBE0 hybrid-functional reference forces that include the exact-exchange contribution.  The resulting \varpro{} force-matching procedure eliminates linear parameters analytically for each trial set of nonlinear parameters and minimizes only the projected residual.  Applied to CP2K reference forces for liquid water, this strategy yields new \qtip-like potentials suitable for efficient classical MD and PIMD.

The approach extends earlier force-matched water models in three respects.  First, it raises the reference level for the production fit from semilocal density-functional forces to PBE0 hybrid-functional forces with explicit exact exchange.  Second, it improves the robustness and interpretability of the optimization by separating amplitudes from nonlinear shape parameters.  Third, it enables a cleaner comparison of short-range O--O interaction forms, showing that a Buckingham repulsion is naturally compatible with the projected optimization and provides a more physically flexible short-range description than an LJ wall.  The fitted potentials reproduce the main first-principles and experimental structural features of liquid water and remain stable in PIMD, making them useful for simulations where direct first-principles PIMD at the hybrid-functional level would be prohibitively expensive.

More broadly, the method is not limited to water.  Any analytic force field in which subsets of parameters enter linearly can benefit from the same projection strategy.  This includes molecular liquids, hydrogen-bonded materials, and hybrid physics-based/data-driven potentials in which linear basis amplitudes coexist with nonlinear descriptor or range parameters.  The present water application therefore provides a transparent benchmark for a general and computationally inexpensive route from accurate first-principles reference data to simulation-ready potentials.

\section*{Author contributions}

J.K. and T.S. generated and analyzed the force-matched water potentials.  K.K. contributed to the path-integral simulation strategy and analysis.  T.D.K. conceived and supervised the project, validated the results, and wrote the manuscript.  All authors discussed the results and contributed to the manuscript.

\section*{Data availability}

The data supporting the findings of this study are available from the corresponding author upon reasonable request.  The reference configurations, fitted parameters, and input files can be provided as supplementary material upon submission.  The classical and path-integral simulations were performed with \texttt{RPMD\_Mainz}; the parameter values required to reproduce the TPSS-D3, PBE, PBE0-LJ, and PBE0-oogam simulations discussed here are given in Table~\ref{tab:pbe0_oogam_parameters}.

\section*{Acknowledgements}

The authors gratefully acknowledge the computing time provided on the high-performance computer Noctua 2 at the NHR Center PC2. This is funded by the Federal Ministry of Education and Research and the state governments participating on the basis of the resolutions of the GWK for national high-performance computing at universities (www.nhr-verein.de/unsere-partner).

\section*{Declarations}

\textbf{Conflict of interest} The authors declare no competing interests.

\end{document}